\documentclass[aps,twocolumn,showpacs,floatfix,superscriptaddress,a4]{revtex4-1}

\usepackage[utf8]{inputenc} 
\usepackage{dcolumn}
\usepackage{graphicx}	
\usepackage{chemformula}
\usepackage{amsfonts}
\usepackage{amsmath,bm,amssymb}	
\usepackage{blkarray}
\usepackage{braket} 
\usepackage{gensymb}
\usepackage{kotex}	
\usepackage{dcolumn}   
\usepackage{color}
\usepackage{ulem}

\def\BigRoman{\uppercase\expandafter{\romannumeral\number\count 255 }}
\def\Romannumeral{\afterassignment\BigRoman\count255=}

\begin{document}

\title[Kiem, Sim, Yoon, and Han]{First-principles-based calculation of branching ratio for 5$\boldsymbol{d}$, 4$\boldsymbol{d}$, and 3$\boldsymbol{d}$ transition metal systems}

\author{Do Hoon Kiem}
\author{Jae-Hoon Sim}
\author{Hongkee Yoon}
\author{Myung Joon Han}

\address{Department of Physics, Korea Advanced Institute of Science and Technology (KAIST), Daejeon 34141, Korea}

\vspace{10pt}


\begin{abstract}
	A new first-principles computation scheme to calculate `branching ratio' has been applied to various $5d$, $4d$, and $3d$ transition metal elements and compounds. This recently suggested method is based on a theory which assumes the atomic core hole interacting barely with valence electrons.	While it provides an efficient way to calculate the experimentally measurable quantity without generating spectrum itself, its reliability and applicability should be carefully examined especially for the light transition metal systems. Here we select 36 different materials and compare the calculation results with experimental data. It is found that our scheme well describes 5$d$ and 4$d$ transition metal systems whereas, for 3$d$ materials, the difference between the calculation and experiment is quite significant. It is attributed to the neglect of core-valence interaction whose energy scale is comparable with the spin-orbit coupling of core $p$ orbitals.	
\end{abstract}
\maketitle

\section{Introduction}
X-ray techniques have been established as one of the most standard tools for physics research and the wide variety of related fields \cite{cowan1981theory,schulke2007electron,de2008core,van2015theory,Rehrtheoretical2000RMP}. Ever since its first discovery of x-ray, tremendous advancements have been made in both experimental facilities and theoretical frameworks. Nowadays its wide energy range and angle resolution as well as the elaborate resonant techniques enable us to study various scientific phenomena. In condensed matter physics, it often serves as a unique tool to detect or identify intriguing phenomena which can hardly be accessed by other techniques \cite{kim_phase-sensitive_2009,kim_magnetic_2012,jeong_direct_2017}. The measured x-ray data often contain lots of information for the material of interest, and its interpretation is highly non-trivial beyond the capability of a simple theoretical model or calculation. The experimental achievement and advancement require to develop new theoretical frameworks and computation schemes \cite{kotani_resonant_2001,ament2011resonant,van2015theory,Rehrtheoretical2000RMP}. Along with this line, the first-principles-based methods have been also quite actively explored \cite{benedetti_ab_2001, prendergast_x-ray_2006,jiang_core-hole_2004, taillefumier_x-ray_2002, schwitalla_electron_1998,laskowski_understanding_2010,PhysRevB.67.024431,shirley_ab_1998,ozaki_prl_2017}.

Recently, we proposed a simple first-principles-based technique to directly calculate the branching ratio (BR) through the formula derived by Thole and van der Laan \cite{thole_linear_1988}. 
This approach enables us to calculate BR and therefore to make a quantitative comparison with experiments without any elaborate heavy computations such as the core hole pseudo-potential generations and Bethe-Salpeter-type many-body calculations \cite{PhysRevB.80.075102,ozaki_prl_2017,PhysRevB.67.024431,shirley_ab_1998,luitz_partial_2001}. 
Not only the BR has a lot of physical meaning itself, but also it is useful to directly compare theory to experiments.
Although our method is based on a crude approximation of atomic theory \cite{thole_linear_1988}, the previous calculations for iridium double perovskites show a good agreement with x-ray absorption spectroscopy (XAS) data \cite{sim_calculating_2016}. Not just because it is simple enough and computationally cheap, but also because the same formula provides the information of spin-orbit coupling (SOC) strength, $\braket{\boldsymbol{L}\cdot \boldsymbol{S}}$, the successful application to iridates is encouraging especially considering recent great interests in large SOC materials \cite{pesin_mott_2010,soumyanarayanan_emergent_2016,hasan_colloquium_2010,qi_topological_2011,kim_novel_2008,kim_phase-sensitive_2009,jackeli_mott_2009,kim_kitaev_2015,chaloupka_kitaev-heisenberg_2010,jeong_direct_2017,kim_magnetic_2012}. 
While its success is largely attributed to the large atomic number of iridium which validates the atomic picture and justifies the ignorance of the core-valence interaction, further investigation of its applicability is an important open issue. If this technique can be reliable for lighter transition metal (TM) systems, it can serve as a useful tool for a wider range of material research. 

In this paper, we investigate the applicability of this technique. We apply this method to various TM elements and compounds. For $5d$ systems, we examined twelve different materials containing Hf, Ta, W, Re, Ir, and Pt. The calculation results are in good agreement with the experimental data. The results of six different $4d$ materials are also in agreement with experiments. For 3$d$ TMs, eight different materials have been considered. It is found that the intensity ratio of $L$ edges is close to the statistical value as expected while the experimental data often exhibit noticeable deviations. This feature can be understood from the sizable core-valence interaction.

\section{Computational Methods}

In the below, we summarize the computational scheme suggested in~\onlinecite{sim_calculating_2016} and the computational details.
The key idea is to focus on correctly estimating BR instead of calculating spectrum itself which requires an elaborate theory to describe the core hole interactions such as Bethe-Salpeter equation and time-dependent density functional theory \cite{albrecht1998ab,rohlfing1998electron,schwitalla1998electron,ankudinov2003dynamic,rehr2010parameter,rehr2000theoretical}. While these advanced methods  can in principle provide the reasonable spectrum for many cases, there is still large room for the methodological improvement from the practical point of view. First, performing these calculations is computationally heavy as briefly mentioned above. Second, these theories have their own formal limitations. For example, the single particle-hole theory is not expected to generate the multiplet structure properly, which is the reason why the previous studies have been mostly focusing on $d^0$ compounds. Finally, each method has its own numerics issues. As a result, if the calculated spectrum is not well compared with the experimental one, empirically it is often difficult to conclude whether it is attributed to the physical reason or simply to the numerical.

\subsection{Branching Ratio Formalism}
To calculate BR, we first calculate $\braket{\boldsymbol{L}\cdot \boldsymbol{S}}$. With our basis set of localized pseudo-atomic orbital (PAO) $\phi_{\alpha,i}$ ($\alpha$: orbital index, $i$: site index) \cite{Openmx}, Kohn-Sham eigenstate is decomposed $\ket{\Psi_{n\boldsymbol{k}}}=\Sigma_{i,\alpha}c_{\alpha,i}^{n,\boldsymbol{k}}\ket{\phi_{\alpha,i}}$ where n and $\boldsymbol{k}$ refer to band index and momentum, respectively. 
The SOC part of Hamiltonian is then estimated as 
\begin{align}
\braket{\boldsymbol{L}\cdot \boldsymbol{S}}&=\sum_{n\boldsymbol{k}}^{\text{occ}}\braket{\Psi_{n\boldsymbol{k}}|\boldsymbol{L}\cdot \boldsymbol{S}|\Psi_{n\boldsymbol{k}}}\nonumber\\
&=\sum_{\epsilon_{n\boldsymbol{k}}}\left( 1.0\times\text{P}_{J=5/2}(\epsilon_{n\boldsymbol{k}})-1.5\times\text{P}_{J=3/2}(\epsilon_{n\boldsymbol{k}})  \right),
\end{align}
where $\text{P}_{J=5/2}$ and $\text{P}_{J=3/2}$ are densities of states for $J=5/2$ and $J=3/2$ state, respectively \cite{sim_calculating_2016}.

In Ref.~\onlinecite{thole_linear_1988}, the core-valence interaction is assumed to be small enough in comparison to the core SOC, and then the core states are well described by total angular momentum quantum number $J$. 
The intensity ratio is
\begin{equation}
{{I_{L_3}}\over{I_{L_2}}}={{2n_h-\braket{\boldsymbol{L}\cdot \boldsymbol{S}}}\over{n_h+\braket{\boldsymbol{L}\cdot \boldsymbol{S}}}}={{2-r}\over{1+r}},
\end{equation}
where $r=\braket{\boldsymbol{L}\cdot \boldsymbol{S}}/n_h$. 

Note that BR can be calculated just from $n_h$ and $\braket{\boldsymbol{L}\cdot \boldsymbol{S}}$. It should also be noted that the correct estimation of $n_h$ can be non-trivial typically due to the hybridization with neighboring atomic orbitals \cite{sim_calculating_2016}, and the value of $n_h$ is dependent on the charge counting method. 
In the current study, we simply take the numerical integration of the partial density of states. 
The deviation ranges caused by this choice are represented by error bars in Fig. 1, 2, and 3.	
	
\subsection{Computational Details}

For the electronic structure calculations, we used `OpenMX' density functional theory software package \cite{Openmx}, which takes the linear combination of numerical PAO as a basis set and the norm-conserving pseudo-potential \cite{ozaki_variationally_2003,ozaki_numerical_2004}.
The cutoff radii for O, Cl, Ti, Cr, Fe, Ni, Cu, Sr, Ru, Rh, Pd, Hf, Ta, W, Re, Ir, Pt, Mn, Co, Sc, Mg, and Ca are 5.0, 7.0, 7.0, 6.0, 6.0, 6.0, 6.0, 10.0, 7.0, 7.0, 7.0, 7.0, 7.0, 7.0, 7.0, 7.0, 7.0, 6.0, 6.0, 7.0, 7.0 and 9.0 a.u., respectively. The Perdew-Burke-Ernzerhof (PBE) exchange-correlation functional has been adopted \cite{perdew_generalized_1996}. For Mott insulators, we used spin-polarized PBE plus $U$ scheme to treat on-site electronic correlations as our main data set. $U_\text{eff}=U-J_\text{H}$ \cite{dudarev_electron-energy-loss_1998,han_2006} values for CuO, RuCl$_3$, IrO$_2$, and Sr$_2$CaIrO$_6$ are 4.0, 1.5, 2.0, and 2.0 eV, respectively \cite{Filippetti_2005CuO,panda_effect_2014,kim_kitaev_2015,persson2005strong,PhysRevB.75.195212,sim_calculating_2016,wang_prb_2006}. We also checked with spin-unpolarized PBE+$U$ considering the recent discussion on this issue \cite{cdft_chen_density_2015,cdft_chen_spin-density_2016,cdft_park_density_2015,cdft_ryee_comparative_2018,cdft_ryee_scirep_2018}. It is found that the calculated values are in a reasonable agreement with each other, and any of our conclusion is not affected by this choice. The SOC was taken into account within a fully relativistic $J$-dependent pseudo-potential scheme in the non-collinear methodology \cite{Openmx}. While the experimental crystal structures are mainly used \cite{DUSSARRAT1998165,douglas_structure_2007,bolzan_structural_1997,kayser_crystal_2014,muller_formation_1968,maeno_superconductivity_1994,johnson_monoclinic_2015,asbrink_refinement_1970,forsyth_effect_1991,adam_crystal_1959,liu_crystal_2007,jones_surface_1997, redman1962cobaltous,johnston1956study}, we also checked the results with varying lattice parameters up to $\pm$3\% (see Sec.3.1).

\begin{figure*}[t] 
	\centering
	\includegraphics[width=2\columnwidth]{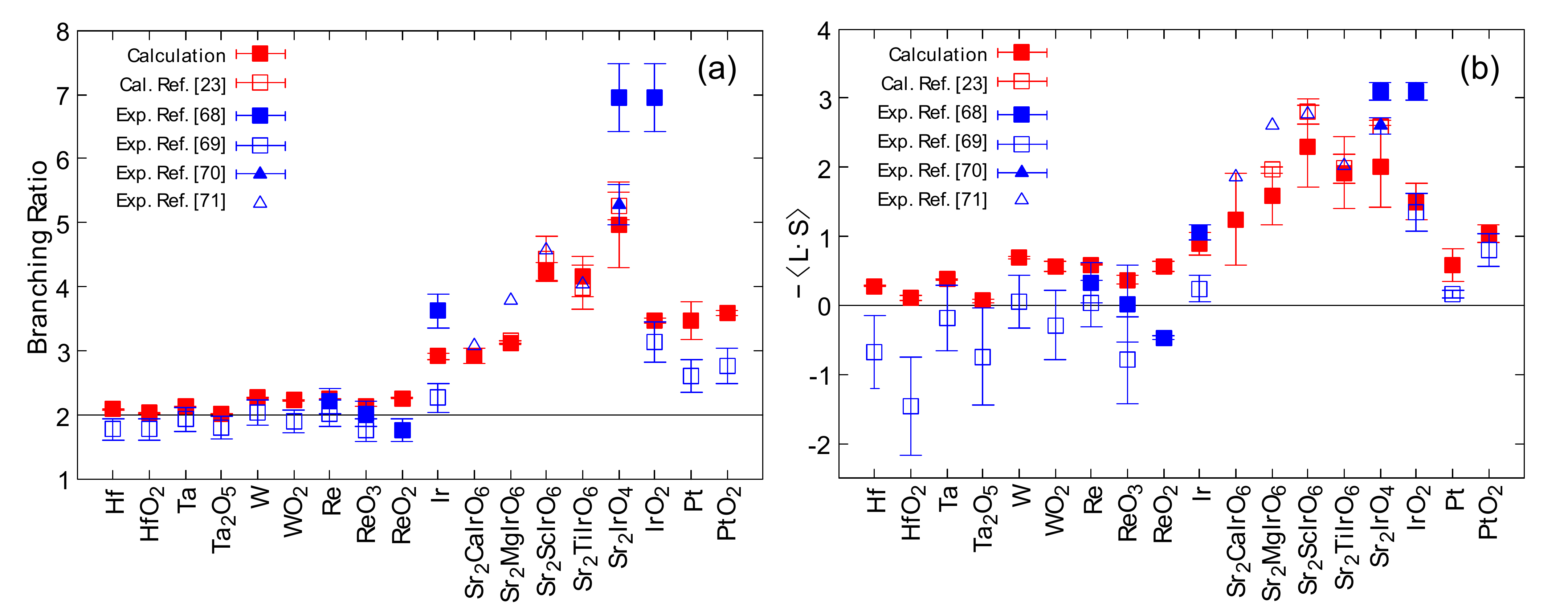}
	\caption{
		The calculation results (red) and experimental data (blue) of (a) BRs and (b) $-\braket{\boldsymbol{L}\cdot \boldsymbol{S}}$ for eighteen different 5$d$ TM systems. Calculation error bars reflect the dependence on the numerical details concerning the $d$-orbital energy range (see main text). The horizontal black lines represent the theoretical value in the absence of spin-orbit interactions (i.e., the statistical value for BR). The calculation results (red) and experimental data (blue) of (a) BRs and (b) $-\braket{\boldsymbol{L}\cdot \boldsymbol{S}}$ for eighteen different 5$d$ TM systems. Calculation error bars reflect the dependence on the numerical details concerning the $d$-orbital energy range (see main text). The horizontal black lines represent the theoretical value in the absence of spin-orbit interactions (i.e., the statistical value for BR). The experimental value of Ref. \onlinecite{chikara_sr_2015} is from Sr$_2$Rh$_{0.05}$Ir$_{0.95}$O$_4$, not stoichiometric Sr$_2$IrO$_4$ (see main text for the related discussion). The experimental $\braket{\boldsymbol{L}\cdot \boldsymbol{S}}$ values are obtained by Eq. (2).}
	\label{Fig:label}
\end{figure*}

\section{Results and Discussion}

\subsection{5d Transition Metal Systems}

First, we apply our method to $5d$ systems. A total of 18 different materials have been investigated; Hf, HfO$_2$, Ta, Ta$_2$O$_5$, W, WO$_2$, Re, ReO$_2$, ReO$_3$, Ir, IrO$_2$, Sr$_2$CaIrO$_6$, Sr$_2$MgIrO$_6$, Sr$_2$TiIrO$_6$, Sr$_2$ScIrO$_6$, Sr$_2$IrO$_4$, Pt, and PtO$_2$. Fig.~1(a) and (b) present the calculation results of BR and $\braket{\boldsymbol{L}\cdot \boldsymbol{S}}$, respectively, in comparison with experimental data \cite{clancy_spin-orbit_2012,cho_x-ray_2012,chikara_sr_2015,laguna-marco_electronic_2015}. The red and blue colors represent the calculation and experimental data, respectively. The error bars for the experimental values are taken directly from references. The overall good agreement between calculations and experiments is clearly noticed.

For elemental Hf, Ta, W, Re, and their oxides, BR is quite close to the statistical value, $I_{L_3}/I_{L_2}=2$ \cite{clancy_spin-orbit_2012,cho_x-ray_2012,qi_l2_1987}, and this feature is well reproduced by our calculation. For elemental Ir, the calculation result is located in between the two experimental values. For Pt and PtO$_2$, the BRs are notably greater than the statistical value in both experiment and calculation. 

The BRs of iridium oxides are significantly larger than 2 which is attributed to the combined effect of charge transfer and crystal fields \cite{cho_x-ray_2012, clancy_spin-orbit_2012}. For IrO$_2$, two experiments report BR $\simeq$ 3.1 \cite{cho_x-ray_2012} and $\simeq$ 6.9 \cite{clancy_spin-orbit_2012}, respectively. Our calculation shows that BR is close to 3.1 in good agreement with the data by Cho et al. \cite{cho_x-ray_2012}. Also for Sr$_2$IrO$_4$, there are two experimental data available; BR $\simeq 7.0$ \cite{clancy_spin-orbit_2012} and $\simeq$ 5.3 \cite{chikara_sr_2015}. Our calculation supports the latter. It is noted that the data of Ref.~\onlinecite{chikara_sr_2015} is taken from 5$\%$ Rh-doped sample while Ref.~\onlinecite{clancy_spin-orbit_2012} and our calculation measure the stoichiometric Sr$_2$IrO$_4$. According to Ref.~\onlinecite{chikara_sr_2015}, BR does not change much as a function of Rh concentration in the small doping regime \cite{chikara_sr_2015}. For Ir double perovskites, Sr$_2${\it A}IrO$_6$, Laguna-Marco et al. recently performed experiments \cite{laguna-marco_electronic_2015}, and three compounds ({\it A}=Mg, Ti, Sc) were calculated in the previous study (also presented in Fig.~1). Our calculation not just reproduces the overall feature of BR for iridium oxides (see the trend from Sr$_2$CaIrO$_6$ to IrO$_2$), but it also gives the quantitative agreement with experiments. It is also noted that, as reported in Ref.~\onlinecite{sim_calculating_2016}, the difference between calculation and experiment can further be reduced (especially for the case of Sr$_2$MgIrO$_6$) by considering the possible oxygen vacancy in the experimental situation.	

\begin{figure}[t]
	\centering
	\includegraphics[width=1\columnwidth]{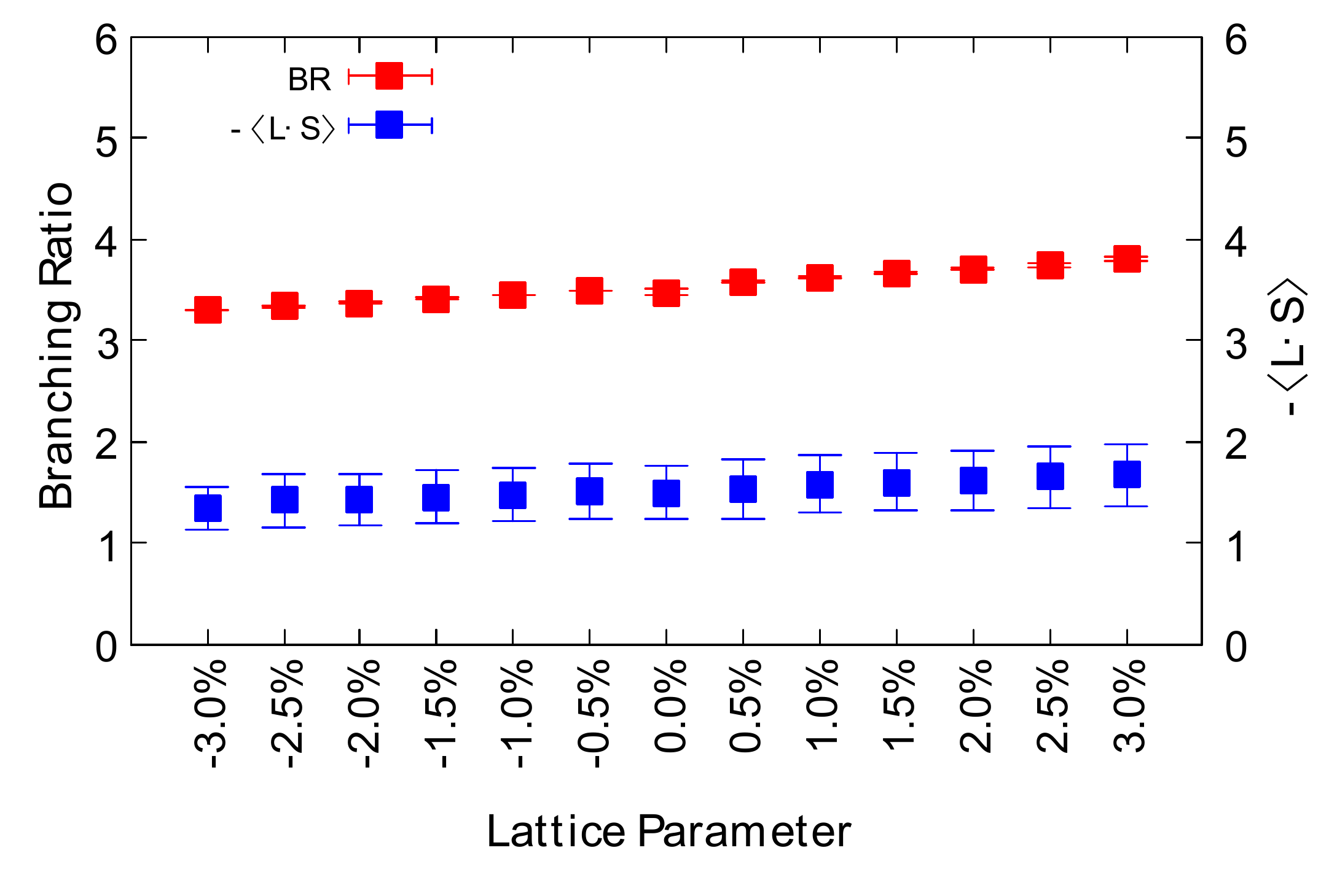}
	\caption{The calculated BR and $-\braket{\boldsymbol{L}\cdot \boldsymbol{S}}$ for IrO$_2$ with varying lattice parameters. The plus and minus sign refers to the enlarged and reduced parameters, respectively, from the experimental value.}
\end{figure}

Fig. 1(b) shows the calculated $\braket{\boldsymbol{L}\cdot \boldsymbol{S}}$ in comparison with experiments. We take the experimental data as in the original papers if $\braket{\boldsymbol{L}\cdot \boldsymbol{S}}$ values are presented. Otherwise, we estimated the experimental values from the nominal charges through $\braket{\boldsymbol{L}\cdot \boldsymbol{S}}=n_h(2-\text{BR})/(1+\text{BR})$.
Note that the measured quantity in the experiment is BR from which $\braket{\boldsymbol{L}\cdot \boldsymbol{S}}$ is estimated, whereas, in our calculation, the more direct quantity is $\braket{\boldsymbol{L}\cdot \boldsymbol{S}}$ as clearly seen in Eq.~(2). Therefore the good agreement between theory and experiment can become a strong indication of the reliability of our method.

As expected, for the materials whose BR is close to the statistical value, $\braket{\boldsymbol{L}\cdot \boldsymbol{S}}$ is close to 0. Overall, the calculation results are in reasonable agreement with experiments. In iridium oxides, $|\braket{\boldsymbol{L}\cdot \boldsymbol{S}}|$ is significantly larger being consistent with Fig.~1(a). It is interesting to note that Sr$_2$IrO$_4$ has a larger BR than Sr$_2$TiIrO$_6$ while both have $d^5$ configuration. 
It is attributed to the smaller crystal field in Sr$_2$IrO$_4$ ($\Delta =$ 3.61 eV) than that of Sr$_2$TiIrO$_6$ ($\Delta =$ 4.19 eV). It is known that $|\braket{\boldsymbol{L}\cdot \boldsymbol{S}}|$ decreases as the crystal field increases \cite{sim_calculating_2016}.

\begin{figure*}[]
	\centering
	\includegraphics[width=2\columnwidth]{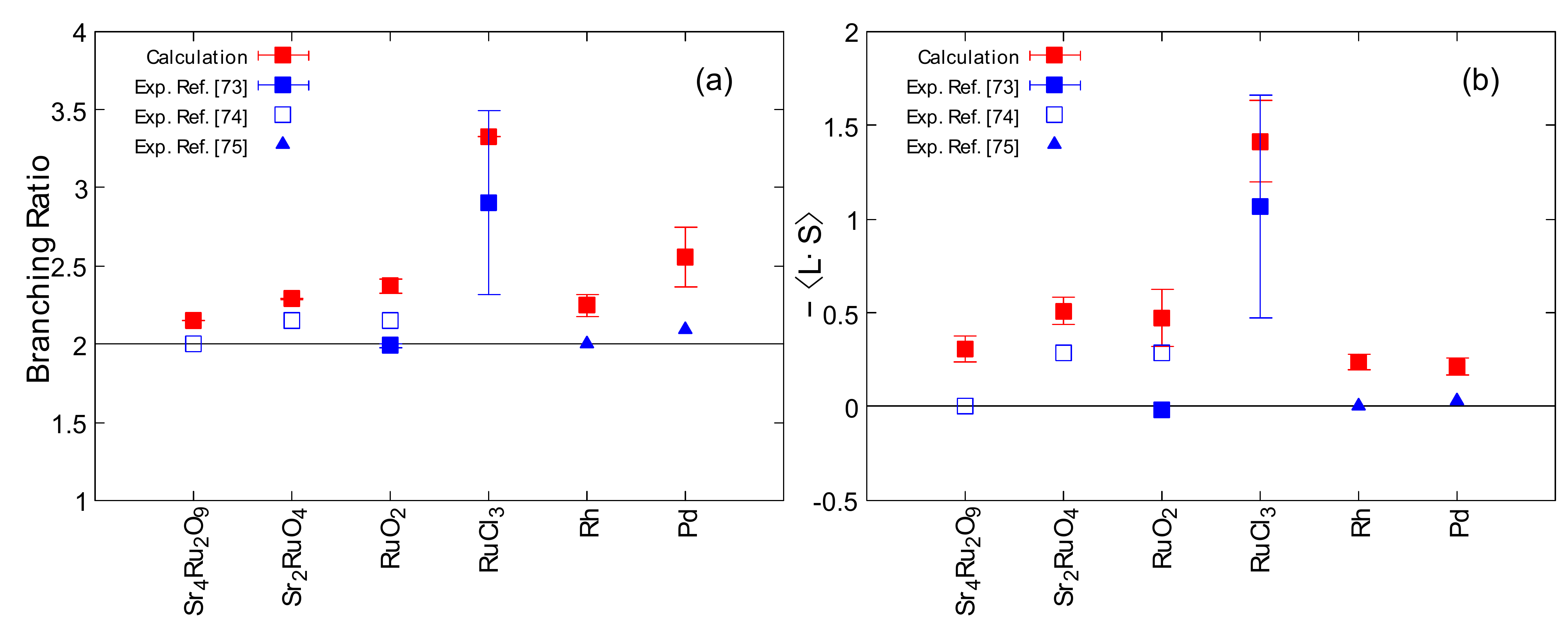}
	\caption{The calculation results (red) and experimental data (blue) of (a) BRs and (b) $-\braket{\boldsymbol{L}\cdot \boldsymbol{S}}$ for six different 4$d$ TM systems. The horizontal black lines represent the statistical value for BR.}
\end{figure*}

Here we emphasize that our calculations consider a wide energy range for $d$-orbital states. In order to represent the ambiguity in taking the $d$-orbital contributions (or in other words, the ambiguity in the projection onto $d$ orbitals) due to the hybridization with O-2$p$ ligands (in the case of oxides) or with other orbitals (in the case of elements), the energy range is set to cover the entire non-negligible PDOS (projected density of states) contributions. It is the reason why the current results are different from the values in Ref.~\onlinecite{sim_calculating_2016} for the case of double perovskites. In the current study, our energy range is typically tens of electron volts while it was set to cover only the main peaks in Ref.~\onlinecite{sim_calculating_2016}. Considering that this type of ambiguity, the good agreement between calculation and experiment supports the reliability of the method. We also check the error range caused by varying lattice parameters. In Fig.~2, the calculation results of IrO$_2$ are presented in which its lattice parameter is changed by $\pm 3\%$. It is found that the dependence is not significant. We also the same feature for other materials.

\subsection{4d Transition Metal Systems}

$4d$ TM systems are of particular interest since our method has only been applied to $5d$ materials and its applicability is expected to be limited as the atomic number decreases. Here we select six different materials for which the experimental data are available; RuO$_2$, Sr$_2$RuO$_4$, Sr$_4$Ru$_2$O$_9$, RuCl$_3$, Rh, and Pd. Fig.~2(a) and (b) shows the calculated BR and $\braket{\boldsymbol{L}\cdot \boldsymbol{S}}$ (red colors), respectively, along with experimental data (blue colors) \cite{plumb_2014,hu_multiplet_2000,sham_l-edge_1985} whose error bars are taken from the original reference papers. The experimental data are fairly well reproduced by our calculations.

For Sr$_4$Ru$_2$O$_9$, Sr$_2$RuO$_4$, RuO$_2$, Rh, and Pd, the experimental BRs are close to the statistical value which indicates that the effect of SOC is not significant as also reflected in Fig. 2(b). The calculation results are in good agreement with experimental data while they tend to overestimate. RuCl$_3$ is of great recent research interest for which the novel Kitaev physics can be realized due to the sizable SOC \cite{kim_crystal_2016,kim_kitaev_2015,plumb_2014,banerjee_neutron_2017,catuneanu_topological_2016}. Indeed, $|\braket{\boldsymbol{L}\cdot \boldsymbol{S}}|$ and BR are much larger in this material, see Fig. 2(a) and (b). The experimental values are in reasonable agreement with our calculation within the error bar ranges. Note that $J=5/2$ state (not $J=3/2$) is solely responsible for so-called $J_{\text{eff}}=1/2$ in this material, which leads to the larger intensity at $L_3$ edge and therefore also to the larger BR. It is also noted that, within the simple ionic picture, Ru ion in this material has $d^5$ configuration whose $|\braket{\boldsymbol{L}\cdot \boldsymbol{S}}|$ value should be smaller than that of $d^4$ according to the naive charge counting. This example clearly shows that the elaborate electronic structure information is important to predict the effect of SOC and the related experimental quantity.

\subsection{3d Transition Metal Systems}

\begin{figure}[t]
	\centering
	\includegraphics[width=1\columnwidth]{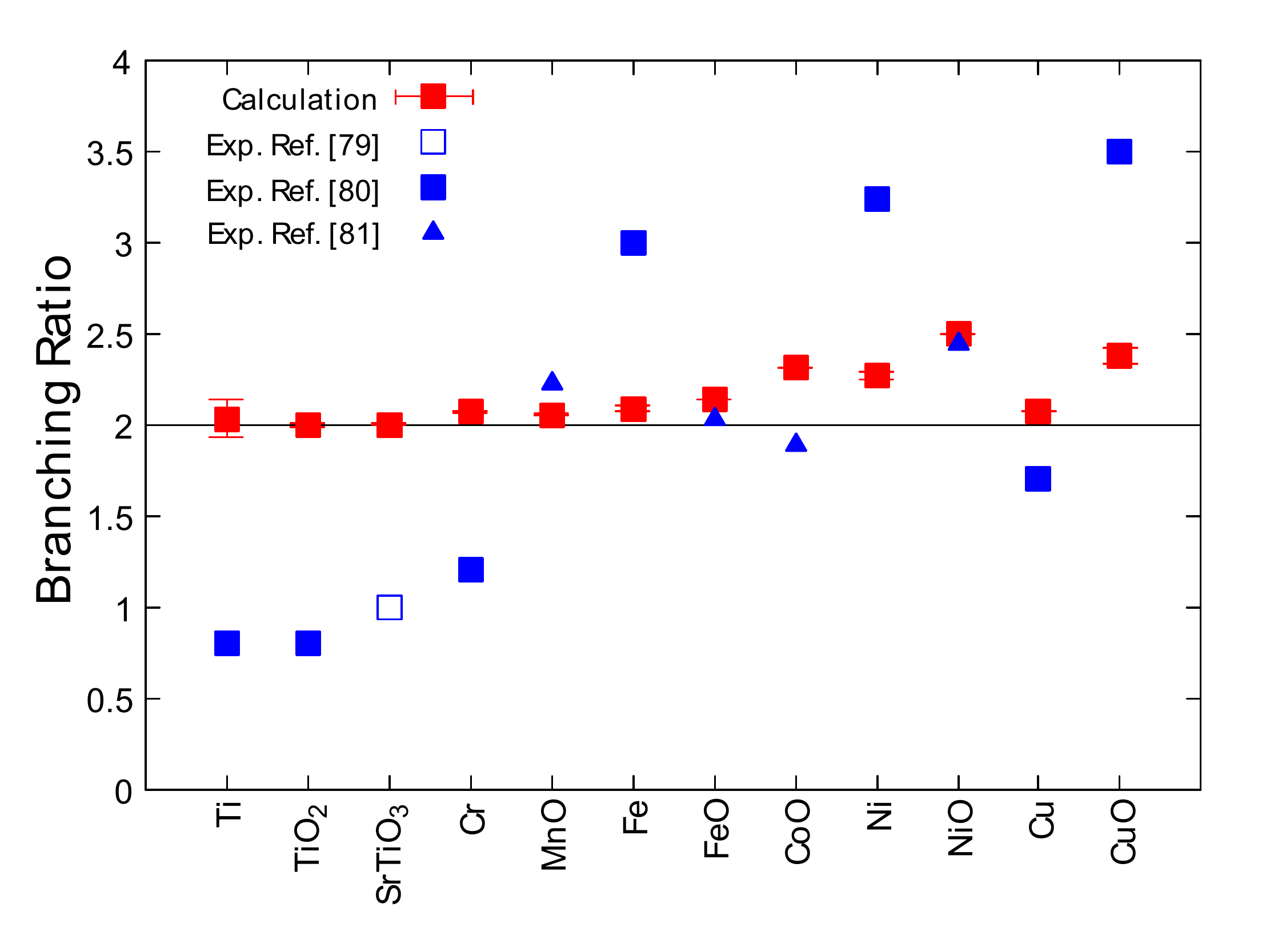}
	\caption{The calculation results (red) and experimental data (blue) of BR for eight different 3$d$ TM systems. The experimental BR of SrTiO$_3$ is an approximate value taken as mentioned in Ref. \cite{van_der_laan_polaronic_1990}. The horizontal black line represents the statistical value for BR. }
\end{figure}

Finally, we investigate 3$d$ TM elements and compounds. A total of twelve systems are taken into account; Ti, TiO$_2$, SrTiO$_3$, Cr, MnO, Fe, FeO, CoO, Ni, NiO, Cu, and CuO. Our calculation results of BR (red colors) are represented in Fig.~3 in comparison with experiments (blue colors) \cite{van_der_laan_polaronic_1990,leapman_study_1982,comment}. Different from the case of 4$d$ and 5$d$ materials, the calculated results significantly differ from the experimental data. 
The calculated $\braket{\boldsymbol{L}\cdot \boldsymbol{S}}$ is small and the BRs are all quite close to the statistical value. It is attributed to the assumption that the core-valence interaction is negligible which can hardly be relevant to light elements. In the case of 3$d$ TMs, the SOC splitting of 2$p_{1/2}$ and $2p_{3/2}$ core states is typically 5 - 20 eV which is comparable with the core-valence interaction roughly a few eV \cite{degroot19902p,haverkort2012multiplet,Luder.PhysRevB.96.245131}. For example, the core-valence interactions are shown to be 5.29--5.6 eV and 6.67--6.8 eV in Mn$^{2+}$ and Ni$^{2+}$, respectively \cite{degroot19902p,haverkort2012multiplet,Luder.PhysRevB.96.245131}.
These values are just slightly increased by considering the core hole states \cite{Luder.PhysRevB.96.245131}.
It is in sharp contrast to the case of 4$d$ and 5$d$ TMs where the SOC of core $2p$ electrons is several hundred eV or even more  \cite{plumb_2014,qi_l2_1987,cho_x-ray_2012,laguna-marco_electronic_2015} whereas the core-valence interaction is typically $\leq$ 5 eV \cite{hu_multiplet_2000,de2008core}.
In order to describe 3$d$ TM systems within the first-principles framework, the more sophisticated techniques are needed to deal with the core holes directly \cite{shirley_bethesalpeter_2005,ogasawara_relativistic_2001,laskowski_understanding_2010}.

\section{Summary}
We investigated various TM elements and compounds by means of a recently-developed first-principles computation scheme. The BR and $\braket{\boldsymbol{L}\cdot \boldsymbol{S}}$ are calculated and systematically compared with experiments. For $4d$ and $5d$ materials with various charge valencies, this computation method gives good agreement with experiments.
For 3$d$ systems, on the other hand, the difference between calculation and experiment becomes significant due to the core-valence interaction neglected in the calculation. The current study establishes the reliability and applicability of this new computation scheme for heavy TM systems, which provides an efficient new way to make a comparison with experiments.

\section{Acknowledgment}
M.J.H thanks Michel van Veenendaal for useful discussion. This work has supported by the National Research Foundation of Korea(NRF) grant funded by the Korea government(MSIT)(No. 2018R1A2B2005204) and Creative Materials Discovery Program through the NRF funded by MSIT (No. 2018M3D1A1059001).


%

\end{document}